\documentclass{article}

\usepackage{arxiv}

\usepackage[utf8]{inputenc} 
\usepackage[T1]{fontenc}    
\usepackage{hyperref}       
\usepackage{url}            
\usepackage{booktabs}       
\usepackage{amsfonts}       
\usepackage{nicefrac}       
\usepackage{microtype}      
\usepackage{lipsum}		
\usepackage{graphicx}
\usepackage{natbib}
\usepackage{doi}
\usepackage{amsmath,amssymb,amsfonts}%
\usepackage{cleveref} 
\usepackage{natbib}
\usepackage{graphicx}%
\usepackage{multirow}%
\usepackage{amsthm}%
\usepackage{mathrsfs}%
\usepackage[figuresright]{rotating}%
\usepackage[title]{appendix}%
\usepackage{xcolor}%
\usepackage{textcomp}%
\usepackage{manyfoot}%
\usepackage{booktabs}%
\usepackage{algorithm}%
\usepackage{algorithmicx}%
\usepackage{algpseudocode}%
\usepackage{program}%
\usepackage{listings}%
\usepackage{mathtools}
\usepackage{caption,subcaption}   

\title{Modeling the mechanosensitive collective migration of cells on the surface and the interior of morphing soft tissues}
\renewcommand{\shorttitle}{Modeling the mechanosensitive collective migration of cells}

\author{
    Jaemin ~Kim \\
	Sibley School of Mechanical and Aerospace Engineering \\
	Cornell University \\
	Ithaca, NY 14853, USA \\
	\And
    Mahmut Selman ~Sakar \\
	Institutes of Mechanical Engineering and Bioengineering \\
        Ecole Polytechnique F\'ed\'erale de Lausanne \\
        Lausanne, Switzerland \\ 
	\And 
    Nikolaos ~Bouklas\thanks{Corresponding author, nb589@cornell.edu} \\
	Sibley School of Mechanical and Aerospace Engineering \\
        Cornell University \\
	Ithaca, NY 14853, USA
}

\begin{document}
\maketitle

\begin{abstract}
Cellular contractility, migration, and extracellular matrix (ECM) mechanics are critical for a wide range of biological processes including embryonic development, wound healing, tissue morphogenesis, and regeneration. Even though the distinct response of cells near the tissue periphery has been previously observed in cell-laden microtissues, including faster kinetics and more prominent cell-ECM interactions, there are currently no models that can fully combine coupled surface and bulk mechanics and kinetics to recapitulate  the morphogenic response of these constructs. Mailand \textit{et al.} (2019) had shown the importance of active elastocapillarity in cell-laden microtissues, but modeling the distinct mechanosensitive migration of cells on the perifery and the interior of highly deforming tissues has not been possible thus fur, especially in the presence of active elastocapillary effects. This paper presents a framework  for understanding the interplay between cellular contractility, migration, and ECM mechanics in dynamically morphing soft tissues accounting for distinct cellular responses in the bulk and the surface of tissues. The major novelty of this approach is that it enables modeling the distinct migratory and contractile response of cells residing on the tissue surface and the bulk, where concurrently the morphing soft tissues undergoes large deformations driven by cell contractility. Additionally, the proposed model is validated through simulation results that capture the changes in shape and cell concentration for wounded and intact microtissues, enabling the interpretation of experimental data. The numerical procedure that accounts for mechanosensitive stress generation, large deformations, diffusive migration in the bulk and a distinct mechanism for diffusive migration on deforming surfaces is inspired from recent work on bulk and surface poroelasticity of hydrogels involving elastocapillary effects, but in this work a two-field weak form is proposed and is able to alleviate numerical instabilities that were observed in the original method that utilized a three-field mixed finite element formulation.
\end{abstract}

\keywords{Morphogenesis \and Mechanotransduction \and Microtissue \and Contractility \and Cell migration \and Wound healing}

\section{Introduction}
The study of the physical interactions between cells and the surrounding extracellular matrix (ECM) is instrumental for diverse fields, including developmental biology, orthopedics, physiotherapy and oncology \citep{chen2007cell,ma2008cell,watt2011cell}. The ability to manipulate and guide these interactions is crucial to engineer implantable tissues for regenerative medicine \citep{schmidt2000acellular,tepole2017computational}. The interactions between cells and ECM involve dynamic feedback loops that coordinate physiological responses to maintain an equilibrium state, which is called ``homeostasis'' \citep{brown1998tensional,humphrey2003continuum,eichinger2021mechanical}. When the homeostatic state is disrupted, cells respond to restore the integrity, architecture, and function of the underlying tissue through a cascade of mechanical events that involve cell migration and matrix remodelling \citep{hinz2012recent,nour2019review,cai2007multi}. 

Remodeling in connective tissues is driven by traction forces applied by fibroblasts residing inside and on the surface of the tissues \citep{grinnell2010cell,mammoto2010,foolen2015shaping,pereira2016mesenchymal,van2018mechanoreciprocity,matis2020mechanical}. Reconstituted collagen and fibrin gels seeded with fibroblasts serve as "tissue equivalents" for the study of mechanical interactions among cells and the surrounding fibrous matrix \citep{bell1979production,stopak1982connective,holle2016vitro,eichinger2021computational,eichinger2021mechanical}. Existing experimental platforms provide limited information on the distribution of stress and strain within the tissue equivalents. Given the complexity of the problem, advanced mathematical models that can interpret experimental data, isolate the factors influencing mechanical behavior at the cellular and tissue levels, and predict biomechanical responses in novel contexts are urgently needed \citep{humphrey2003continuum,eichinger2021computational}. These mathematical models can also facilitate hypothesis-driven research \citep{kim2023model}.

Tissue remodelling is a multi-scale process, and hence, various models have been developed to capture the behavior of cells with varying details, including molecular dynamics (MD) simulations, agent-based models, and continuum models with phenomenological interaction rules \citep{shaebani2020computational}. Although MD simulations and agent-based modeling offer a high degree of spatial resolution, they become impractical for studying large-scale tissue mechanics involving many cells \citep{wang2020continuum}. To overcome this limitation, continuum approaches that model locally averaged details of the cell/ECM dynamics have been proposed \citep{byrne2009individual}. A number of continuum models have been developed to investigate how cellular contractility and mechanosensing affect tissue deformation. Certain models consider the effects of ECM alignment and dynamic cross-linking of ECM proteins \citep{legant2009microfabricated,deshpande2006bio,vernerey2011constrained,shenoy2016chemo,tepole2017computational,ban2018mechanisms,gonzalez2018mechanical,baker2015cell,abhilash2014remodeling} while other models captured the effects of cell migration \citep{gonzalez2018mechanical,banerjee2019continuum,kim2020model,mailand2021tissue,kim2023model}. None of these models have taken into account the distinct responses of cells residing on the surface and in the core of the fibrous tissue, despite growing experimental evidence suggesting a connection between cell migration and the geometric and mechanical properties of the tissue \citep{legant2012force,sunyer2016collective,grolman2020extracellular}. Indeed, cells residing on the surface of a tissue experience a very different physical environment than the cells surrounded by a network of dense matrix.

In our previous work, we demonstrated that surface stresses arise in constrained fibroblast-populated collagen gels, leading to the morphogenesis of fibrous microtissues \citep{mailand2019surface}. We introduced computational models that incorporate both surface and bulk contractile stresses with the passive elasticity of the ECM in 2D \citep{mailand2019surface} and 3D \citep{kim2020model}. More recently, we updated the model to account for the interplay between cell migration, contractility, and ECM mechanics in a dynamically morphing soft tissue \citep{kim2023model}. In that framework, cell migration occurs through a combination of active cell migration and passive tissue deformation. As cells generate contractile stresses, the tissue contracts and deforms. If the tissue is mechanically constrained, an inhomogeneous stress field could arise. Cells are modeled as mechanosensitive in a way that they actively move to regions where they can apply higher forces. An important limitation in \citep{kim2020model}, is that the surface cell concentration is directly tied to the bulk cell concentration, and cells cannot move in a different modality on the surface or in the bulk, and there can also be no exchange between the two species populations. To further advance our understanding of the interplay between cellular mechanics and tissue morphogenesis, it is instrumental to develop a continuum mechanics framework that accounts for distinct surface cell kinetics on deforming soft bodies. 

The theoretical foundation of this formulation has been explored in few recent publications. \citet{mcbride2011geometrically} presented a nonlinear theory to account for species migration in both the bulk and on surfaces but did not present a computational framework. \citet{lucantonio2016continuum} further specialized this theory for thermo-responsive hydrogels and membranes and applied it to drug delivery systems, and also presented a mixed finite element formulation for the problem. A more recent study by \citet{kim2023finite} has extended this framework for hydrogels in lengthscales where surface effects (e.g. surface tension) could be important and drive deformation and species migration in the bulk and on the free surfaces of the materials, leading to complex concentration profiles at equilibrium; where a mixed finite element formulation was also presented. These advancements open up exciting possibilities for investigating complex biological systems, such as cell-ECM systems  where collective migration and contraction of cells is distinct on the tissue periphery and in the bulk, and ``active'' surface effects become prominent.

Here, we present a continuum mechanics framework that can account for both surface and bulk cell migration and tissue contractility in dynamically morphing soft tissues, along with a corresponding finite element implementation in a total Lagrangian setting. The proposed model allows for cells to migrate in the bulk, on the deforming tissue surfaces, and also transition between the bulk and the surface while allowing for cell/ECM interactions. The model incorporates the coupling of cell concentration and local deformation state with both the associated contractile stress generation and cell migration. In \Cref{sec:A nonlinear theory}, we summarize the proposed continuum theory. \Cref{sec:Specific consideration for cell-ECM interaction} propose the novel free energy for mechanosensitive cell migration and contractility. \Cref{sec:Mixed finite element formulation} presents the finite element formulation and \Cref{sec:Results and discussion} presents simulations of morphing microtissues and wound closure.

\section{A nonlinear theory}\label{sec:A nonlinear theory} 
We present a novel phenomenological continuum framework that accounts for the distinct response of mesenchymal cells contracting and migrating in extracellular matrix (ECM) to recapitulate tissue-level mechanics. Our theory accounts the temporal coupling between cell migration and contractility, resulting in an equilibrium homeostatic state that describes inhomogeneous 3D deformation and cell concentration. In our notation, $\{\bullet\}$ and $\{\widetilde{\bullet}\}$ denote bulk and surface quantities, respectively, for a body occupying a volume $\mathnormal{V}$ bounded by an outer surface denoted as $\mathnormal{S}$. It is important to note that a surface quantity is not equivalent to the bulk quantity evaluated on the surface. The general notation convention and nomenclature follow \citep{kim2023finite}, where the reader can also follow some of the key concepts in differential geometry necessary for the description of the kinematics of deforming surfaces.

\begin{figure}[!ht]
    \centering
    \includegraphics[width=0.6\linewidth]{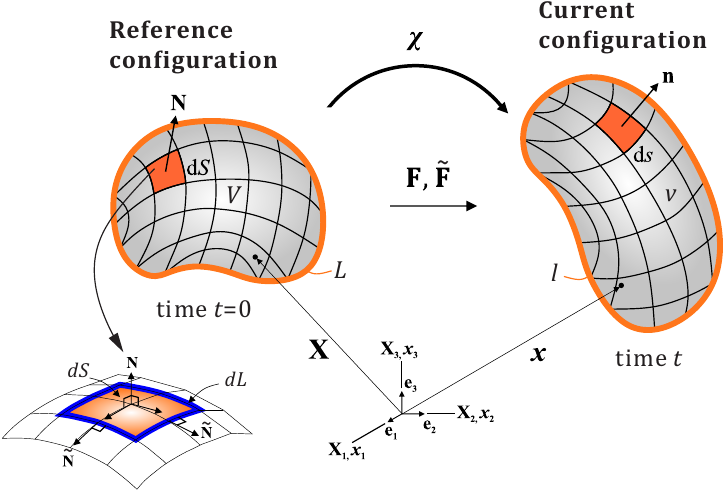}
    \caption{Schematic illustration of the reference and current state of a continuum body. The reference volume and surface, and boundary are denoted by $\mathnormal{V}$ and $\mathnormal{S}$, and $\mathnormal{L}$ respectively. The normal vector to the surface in the reference and current configuration ($\mathbf{N}$ and $\mathbf{n}$) and the bi-normal vector to the boundary ($\widetilde{\mathbf{N}}$) are shown, where the over-tilde indicates surface quantities.}
    \label{fig:Fig1}
\end{figure}

\subsection{Kinematics}\label{sec:Kinematics}
Let $\mathnormal{V}$ be a fixed reference configuration of a continuum body $\mathcal{B}$. We use the notation $\chi : \mathnormal{V} \rightarrow \mathbb{R}^3$ for the deformation of body $\mathcal{B}$. A motion $\chi$ is the vector field of the mapping $\textit{\textbf{x}} = \chi(\textbf{X})$, of a material point in the reference configuration $\textbf{X} \in \mathnormal{V}$ to a position in the deformed configuration $\textit{\textbf{x}} \in \mathnormal{v}$. The kinematics of a typical particle are described by the displacement vector field in the spatial description, $\textbf{u}(\textbf{X},t)=\textit{\textbf{x}}\,(\textbf{X},t)-\textbf{X}$. The kinematics of an infinitesimal bulk element are described by
\begin{subequations}\label{eqn:deformation_gradient_bulk}
\begin{align}
    \textbf{F}(\textbf{X},\textit{t})&=\frac{\partial\chi(\textbf{X},\textit{t})}{\partial\textbf{X}}=\boldsymbol{\nabla}_{ \textbf{X}}\,\textit{\textbf{x}}\,(\mathbf{X},t)\label{eqn:forward_deformation_gradient_bulk}\\
    \textbf{F}^{-1}(\textit{\textbf{x}},\textit{t})&=\frac{\partial\chi^{-1}(\textit{\textbf{x}},\textit{t})}{\partial\textit{\textbf{x}}}=\boldsymbol{\nabla}_{ \textit{\textbf{x}}}\,\textbf{X}(\textit{\textbf{x}},\textit{t})\label{eqn:inverse_deformation_gradient_bulk}
\end{align}
\end{subequations}
\noindent where $\textbf{F}(\textbf{X},\textit{t})$ and $\textbf{F}^{-1}(\textit{\textbf{x}},\textit{t})$ are the deformation gradient and inverse deformation gradient, respectively. Note that $\mathnormal{J}(\textbf{X},\textit{t})=\textrm{d}\textit{v}/\textrm{d}\textit{V}=\text{det} \, \textbf{F}(\textbf{X},\textit{t})>0$ is the Jacobian determinant defining the ratio of a volume element between the material and spatial configuration.

The surface displacement $\widetilde{\textbf{u}}(\widetilde{\textbf{X}},t)$ can be determined by $\mathbf{u}(\mathbf{X},t)\vert_{\mathnormal{S}} = \widetilde{\textbf{u}}(\widetilde{\textbf{X}},t)$, which imposes kinematic conformity between the bulk and the surface. The motion of an arbitrary differential vector element connecting two material points in the bulk can be mapped by the deformation gradient $\textbf{F}$ through the motion of the body. However, a unit normal vector $\textbf{N}$ in the material configuration cannot be transformed into a unit normal vector $\textbf{n}$ in the spatial configuration via the deformation gradient, as shown in \Cref{fig:Fig1} \citep{holzapfel2000nonlinear,steinmann2008boundary}. This motivates us to follow the kinematics of an infinitesimal surface element \citep{steinmann2008boundary,kim2020model}. 
\begin{subequations}\label{eqn:deformation_gradient_surf}
\begin{align}
    \widetilde{\textbf{F}}(\widetilde{\textbf{X}},\textit{t})&
    =\frac{\partial\chi(\widetilde{\textbf{X}},\textit{t})}{\partial\widetilde{\textbf{X}}}\cdot\widetilde{\mathbf{I}}
    =\widetilde{\boldsymbol{\nabla}}_{ \widetilde{\textbf{X}}}\,\widetilde{\textit{\textbf{x}}}\,(\widetilde{\mathbf{X}},t)\label{eqn:forward_deformation_gradient_surf}\\
    \widetilde{\textbf{F}}^{-1}(\widetilde{\textit{\textbf{x}}},\textit{t})&
    =\frac{\partial\chi^{-1}(\widetilde{\textit{\textbf{x}}},\textit{t})}{\partial\widetilde{\textit{\textbf{x}}}}\cdot\widetilde{\textit{\textbf{i}}}
    =\widetilde{\boldsymbol{\nabla}}_{\widetilde{ \textit{\textbf{x}}}}\,\widetilde{\textbf{X}}(\widetilde{\textit{\textbf{x}}},\textit{t})\label{eqn:inverse_deformation_gradient_surf}
\end{align}
\end{subequations}
where $\widetilde{\textbf{F}} (\widetilde{\textbf{X}}, \textit{t})$ and $\widetilde{\textbf{F}}^{-1} (\widetilde{\textit{\textbf{x}}}, \textit{t})$ are the surface deformation gradient and inverse surface deformation gradient. The $\widetilde{\textbf{I}}=\textbf{I}-\textbf{N}\otimes\textbf{N}$ and $\widetilde{\textit{\textbf{i}}}=\textit{\textbf{i}}-\textbf{n}\otimes\textbf{n}$ are the mixed surface unit tensors with the outward unit normal vectors $\textbf{N}$ and $\textbf{n}$, where $\widetilde{\mathbf{I}}$ and $\widetilde{\textit{\textbf{i}}}$ act as a surface (idempotent) projection tensors in material and spatial configurations, respectively. Note that $\widetilde{\mathnormal{J}}(\widetilde{\textbf{X}},\textit{t})=\textrm{d}\textit{a}/\textrm{d}\textit{A}=\text{det} \, \widetilde{\textbf{F}}(\widetilde{\textbf{X}},\textit{t})>0$ is the surface Jacobian determinant defining the ratio of a surface element between material and spatial configuration.

The divergence theorems for bulk and surface are defined by \citep{green1992theoretical,steinmann2008boundary} 
\begin{subequations}\label{eqn:Divergence_Thm}
\begin{align}
    \int_{\mathnormal{V}}\boldsymbol{\nabla}_{ \textbf{X}}\cdot\{\bullet\small\}\,\mathrm{d}\mathnormal{V}
    &=\int_{\mathnormal{S}}\{\bullet\}\cdot\textbf{N}\,\text{d}\mathnormal{S} \label{eqn:Divergence_Thm_Bulk} \\
    \int_{\mathnormal{S}}\widetilde{\boldsymbol{\nabla}}_{ \widetilde{\textbf{X}}}\cdot\{\widetilde{\bullet}\}\,\text{d}S
    &=\int_{\mathnormal{L}}\small\{\widetilde{\bullet}\small\}\cdot\widetilde{\textbf{N}}\,\text{d}\mathnormal{L}
    -\int_{\mathnormal{S}}\widetilde{\kappa}\small\{\widetilde{\bullet}\small\}\cdot\textbf{N}\,\text{d}S \label{eqn:Divergence_Thm_Surf}
\end{align}
\end{subequations}
\noindent where $\widetilde{\textbf{N}}$ is the unit outward bi-normal vectors to the boundary curve, $\mathnormal{L}$, and $\widetilde{\kappa}=-\widetilde{\boldsymbol{\nabla}}_{ \widetilde{\textbf{X}}}\cdot\textbf{N}$ is total curvature (twice the mean surface curvature). 

We decompose the deformation gradient into a volumetric and an isochoric part, $\textbf{F}=(J^{1/3}\textbf{I}) \, \overline{\textbf{F}}$, where $J^{1/3}\textbf{I}$ and $\overline{\textbf{F}}$ are associated with volumetric and isochoric deformation. Following, we utilize $\{\overline{\bullet}\}$ to denote quantities associated with the isochoric part of the deformation gradient. We introduce the strain measures as follows:
\begin{subequations}
\begin{align}
    \mathbf{C} = \mathbf{F}^{\mathrm{T}}\mathbf{F} \quad &\text{and} \quad \mathnormal{I}_{1} = \text{tr}(\mathbf{C}) \\ \overline{\textbf{C}} =\overline{\textbf{F}}^{\text{T}} \, \overline{\textbf{F}} \quad &\text{and} \quad \overline{\mathnormal{I}}_{1} = \text{tr}(\overline{\mathbf{C}})
\end{align}
\end{subequations}
\noindent where $\mathbf{C}$ and $\mathnormal{I}_{1}$ are the right Cauchy-Green tensor and its first principal invariant, and $\overline{\mathbf{C}}$ and $\overline{\mathnormal{I}}_{1}$ are the modified right Cauchy-Green tensor and its first principal invariant.

We introduce the right Cauchy-Green tensors on the surface as
\begin{equation}\label{eqn:Right_Cauchy_Green_tensor_surf}
    \widetilde{\textbf{C}} = \widetilde{\textbf{F}}^{\text{T}} \, \widetilde{\textbf{F}}
\end{equation}
Note that we cannot directly obtain the inverse of the surface right Cauchy-Green tensor due to the fact that it is not full rank (a characteristic that it inherits from the surface deformation gradient tensor). Nevertheless, we can still obtain its inverse form in the generalized sense,
\begin{equation}\label{eqn:Right_Cauchy_Green_tensor_surf_inverse}
   \widetilde{\mathbf{C}}^{-1} = \widetilde{\mathbf{I}}\mathbf{C}^{-1}\widetilde{\mathbf{I}}
\end{equation}
which will be utilized in the forthcoming developments for defining the surface kinetic law. The detailed derivation for the surface kinematics can be found in \citet{green1992theoretical,steinmann2008boundary,do2016differential}.

\subsection{Equilibrium}\label{sec:Equilibrium}
Mechanical equilibrium is assumed to be maintained at all time during the transient process neglecting inertial terms and body forces.
\begin{subequations}\label{eqn:Equilibrium}
\begin{alignat}{2}
    \boldsymbol{\nabla}_{\textbf{X}}\cdot\mathbf{P} + \mathbf{B} &= 0 \quad  &&\text{in} \quad \mathnormal{V} \label{eqn:EquilibriumBulk} \\
    \mathbf{P}\mathbf{N} - \widetilde{\boldsymbol{\nabla}}_{\widetilde{\textbf{X}}}\cdot\widetilde{\mathbf{P}} &= \mathbf{T} \quad  &&\text{on} \quad \mathnormal{S}_{\mathrm{T}} \label{eqn:EquilibriumSurf}\\
    \mathbf{u} &= \mathbf{u}_{p} \quad &&\text{on} \quad \mathnormal{S}_{\mathrm{u}} \label{eqn:EqulibriumDirchletBC}\\
    [[\widetilde{\mathbf{P}}\widetilde{\mathbf{N}}]] &= 0 \quad  &&\text{on} \quad \mathnormal{L} \label{eqn:EqulibriumNeumannBC}
\end{alignat}
\end{subequations}
where $\mathbf{P}$ and $\widetilde{\mathbf{P}}$ are the first Piola-Kirchoff (PK1) stress in bulk and on surface, $\mathbf{B}$ and $\mathbf{T}$ are the body force and the traction vector, $\mathbf{u}$ is the displacement, and $\mathbf{u}_{p}$ is the prescribed displacement on the Dirichlet part of the boundary condition. Note that a Neumann-type boundary condition is also
defined on boundary curves that $[[\bullet]]$ indicates summation over surfaces intersecting on boundary curves \citep{steinmann2008boundary}.

\subsection{Cell balance law}\label{sec:Cell balance law}
Through species balance, the equations for the rate of change of nominal cell concentration, $\dot{\mathnormal{C}}$ and $\dot{\widetilde{\mathnormal{C}}}$ are obtained \citep{mcbride2011geometrically}:
\begin{subequations}\label{eqn:BalanceLaw}
\begin{alignat}{2}
    \dot{\mathnormal{C}} + \boldsymbol{\nabla}_{\textbf{X}} \cdot \mathbf{J} &= \mathnormal{r}
    \quad &&\text{in} \quad \mathnormal{V} \label{eqn:BalanceLawBulk}\\
    \dot{\widetilde{\mathnormal{C}}} + \widetilde{\boldsymbol{\nabla}}_{\textbf{X}} \cdot \widetilde{\mathbf{J}}-\mathbf{J}\cdot\mathbf{N} &= \mathnormal{i}
    \quad &&\text{on} \quad \mathnormal{S}_{\widetilde{\mathnormal{C}}} \label{eqn:BalanceLawSurf}\\
    \widetilde{\mathbf{J}} &= \widetilde{\mathbf{J}}_{p} \quad &&\text{on} \quad \mathnormal{S}_{\widetilde{\mathbf{J}}} \label{eqn:BalanceLawDirchletBC}\\
    [[\widetilde{\mathbf{J}} \cdot \widetilde{\mathbf{N}}]] &= 0
    \quad &&\text{on} \quad \mathnormal{L} 
\end{alignat}
\end{subequations}
where $\dot{\mathnormal{C}}$ and $\dot{\widetilde{\mathnormal{C}}}$ are the rate of change of nominal species concentration in the bulk and on the surface, $\mathbf{J}$ and $\widetilde{\mathbf{J}}$ indicate the nominal flux of species in the reference configuration. The $\mathnormal{r}$ and $\mathnormal{i}$ are the source/sink terms for the number of cells injected into the reference volume and area per unit time, which can be attributed to proliferation/apoptosis. 
$\widetilde{\mathbf{J}}_{p}$ is the prescribed surface flux. The \Cref{eqn:BalanceLaw} are supplemented with initial and boundary conditions
\begin{subequations}\label{eqn:InitialCondition}
\begin{alignat}{2}
    \mathnormal{C} &= \mathnormal{C}_{0} \quad &&\text{at} \quad t=0 \\
    \widetilde{\mathnormal{C}} &= \widetilde{\mathnormal{C}}_{0} \quad &&\text{at} \quad t=0
\end{alignat}
\end{subequations}
where $\mathnormal{C}_{0}$ and $\widetilde{\mathnormal{C}}_{0}$ are the initial nominal species concentration at time $\mathnormal{t}=0$. It is important to note that we neglect cell proliferation/apoptosis in the bulk $\mathnormal{r}=0$ and on the surface $\mathnormal{i}=0$ for this work, which could be very important as we study other systems.

\subsection{Constitutive relations}\label{sec:Constitutive relations}
It has been observed that cells located near the tissue boundary exhibit significantly more polarization than those in the bulk \citep{mailand2019surface,kim2020model}. Hence, besides the free energy density in the bulk, the free energy density on the surface must also be taken into account. We will use the terms ``bulk'' and ``surface'' to denote the respective free energy densities.
\begin{equation}\label{eqn:StrainEnergyFunction}
    \Psi(\mathbf{F},\mathnormal{C},\eta) \quad \text{and} \quad \widetilde{\Psi}(\widetilde{\mathbf{F}},\widetilde{\mathnormal{C}},\widetilde{\eta})
\end{equation}

Following from \Cref{sec:Thermodynamics considerations}, the parentheses of the first six terms in \Cref{eqn:DissipationInequalityChainRule} must be zero to satisfy the inequality for any arbitrary $\dot{\mathbf{F}}$, $\dot{\widetilde{\mathbf{F}}}$, $\dot{\mathnormal{C}}$ and $\dot{\widetilde{\mathnormal{C}}}$, which results in the constitutive relations as follows:
\begin{subequations}\label{eqn:ConstitutiveRelation}
\begin{alignat}{3}
    \mathbf{P} &= \frac{\partial\Psi}{\partial\mathbf{F}} 
    ,\quad
    \mu &&= \frac{\partial\Psi}{\partial\mathnormal{C}}
    ,\quad
    \xi &&= \frac{\partial\Psi}{\partial\eta}
    \label{eqn:ConstitutiveRelationBulk}
    \\
    \widetilde{\mathbf{P}} &= \frac{\partial\widetilde{\Psi}}{\partial\widetilde{\mathbf{F}}}
    ,\quad
    \widetilde{\mu} &&= \frac{\partial\widetilde{\Psi}}{\partial\widetilde{\mathnormal{C}}} 
    ,\quad
    \widetilde{\xi} &&= \frac{\partial\widetilde{\Psi}}{\partial\widetilde{\eta}}
    \label{eqn:ConstitutiveRelationSurf}
\end{alignat}
\end{subequations}
where $\mathbf{P}$ and $\widetilde{\mathbf{P}}$ represent the first Piola-Kirchhoff (PK) stress tensors in the bulk and on the surface, respectively. The variables $\mu$ and $\widetilde{\mu}$ denote the bio-chemical potential in the bulk and on the surface, respectively. The microscopic forces generated by cells and applied to the ECM network are represented by $\xi$ in the bulk and $\widetilde{\xi}$ on the surface. It is important to note that the microscopic forces indicate that an increase in the activation level results in an increase in the force exerted on the ECM network. 

For the seventh term in \Cref{eqn:DissipationInequalityChainRule}, the most obvious way to guarantee the dissipation inequality on surface is to impose the condition \citep{mcbride2011geometrically},
\begin{equation}\label{eqn:bio-chemicalSlavery}
    \mu(\mathbf{X},t) = \widetilde{\mu}(\widetilde{\textbf{X}},t)
    \quad \text{on} \quad \mathnormal{S}
\end{equation}
which is the conformity of the bio-chemical potential between the surface and bulk.

The eighth and ninth terms in \Cref{eqn:DissipationInequalityChainRule} remain negative, respectively, and we adopt a kinetic law \citep{gurtin2010mechanics} to describe the consistent species diffusion that is driven by the bio-chemical potential. The kinetic law is given by:
\begin{subequations}\label{eqn:KineticLaw}
\begin{alignat}{3}
    \mathbf{J}&=-\mathbf{M}\boldsymbol{\nabla}_{\mathbf{X}}\mu 
    \quad \text{with} \quad
    \mathbf{M} = \mathnormal{C}\mathnormal{D}\mathbf{C}^{-1} \label{eqn:KineticLawBulk}
    \\
    \widetilde{\mathbf{J}}&=-\widetilde{\mathbf{M}}\widetilde{\boldsymbol{\nabla}}_{\widetilde{\textbf{X}}}\widetilde{\mu}
    \quad \text{with} \quad
    \widetilde{\mathbf{M}} = \widetilde{\mathnormal{C}}\widetilde{\mathnormal{D}}\widetilde{\mathbf{C}}^{-1} 
    \label{eqn:KineticLawSurf}
\end{alignat}
\end{subequations}
Here, $\mathbf{M}$ and $\widetilde{\mathbf{M}}$ are positive-semidefinite mobility tensors in the bulk and on the surface, respectively. The effective diffusivity of the cells $\mathnormal{D}$ in the bulk and $\widetilde{\mathnormal{D}}$ on the surface are assumed to be isotropic and independent of deformation and concentration as the simplest approximation \citep{bouklas2012swelling}. The kinetic law describes the consistent migration of cells driven by the bio-chemical potential $\mu$. Note that Fick's law for diffusion in the current configuration with a mobility constant $\mathnormal{M}$ is pulled back to the reference configuration to derive our model.

\section{Specific considerations for cell-laden microtissues}\label{sec:Specific consideration for cell-ECM interaction}
In this work, we will focus on a coupled framework for cell-ECM interaction of cell-laden contractile microtissues that considers bulk and surface ``active'' mechanics and kinetics. We have to specialize our choices for the surface and bulk free energy densities, corresponding constitutive laws, and definition of mobility tensors for our theory to be complete and to be able to proceed to the development of the numerical solution scheme. 

\subsection{A specific free energy }\label{sec:A specific free energy}
The free energy in the bulk should account for both passive and active contributions. The passive term captures the elasticity of the collagen network, which is typically considered nearly incompressible \citep{kim2020model}. However, we allow for compressibility to accommodate tissue-level compaction generated by cell contractility. The active term captures the contractile action of fibroblasts residing in the bulk. On the other hand, the free energy on the surface only considers the active contribution, which captures the contractile action of fibroblasts on the surface and leads to a fluid-like response.
\begin{equation}\label{eqn:Strain_energy}
    \Psi(\mathbf{F},\mathnormal{C}) 
    = \Psi_{p}(\mathbf{F}) 
    + \Psi_{a}(\mathbf{F},\mathnormal{C})
    \quad \text{and} \quad
    \widetilde{\Psi}_{a}(\widetilde{\mathbf{F}},\widetilde{\mathnormal{C}})
\end{equation}
where the free energy density on the surface, $\widetilde{\Psi}$, only considers the active contribution capturing the contractile action of fibroblasts on the surface, while the bulk free energy density $\Psi$ is decomposed into the passive component $\Psi_{p}$ due to elastic deformations of the ECM and the active component $\Psi_{a}$ due to cell migration and contraction in the bulk. The active contribution on the surface is captured by $\widetilde{\Psi}_{a}$, which takes into account the distinct mechanism of cell migration and contraction on the surface.

\begin{figure*}[!ht]
    \centering
    \includegraphics[width=0.4\linewidth]{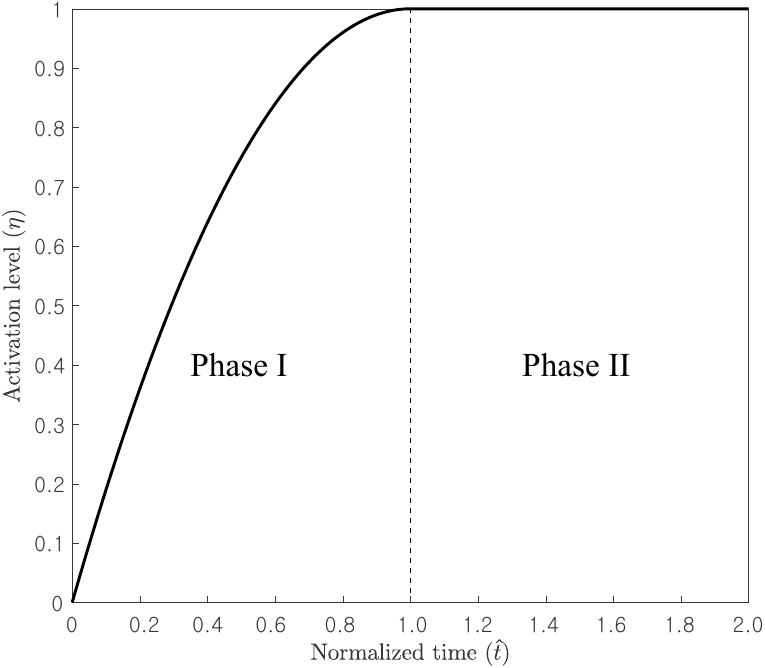}
    \caption{Activation level can map to two characteristic phases, a rapid increase (phase I) followed by a near-homeostatic state (phase II).}
    \label{fig:Fig2}
\end{figure*}

For the ECM response, we adopt a compressible neo-Hookean model \citep{holzapfel2000nonlinear}, 
\begin{equation}\label{eqn:Psi_bulk_passive}
     \Psi_{p}(\mathbf{F}) = \frac{\mathnormal{G}}{2}\left(\overline{\mathnormal{I}}_{1}-3\right) + \frac{\mathnormal{K}}{2}\left(\mathnormal{J}-1\right)^{2}
\end{equation}
where $\mathnormal{G}$ and $\mathnormal{K}$ are the shear and bulk moduli, which can be uniquely defined by the Young's modulus $\mathnormal{E}$ and Poisson's ratio $\nu$. In order to capture the mechano-sensitive coupling that drives the competition of cell migration and contractility, we need to take into account the cell concentration and the level of activation of cell contraction, as well as the deformation. To this end, we propose a set of active free energy functions for the bulk and surface, which aim to capture the distinct mechanisms of cell migration and contraction.
\begin{subequations}\label{eqn:Psi_active}
\begin{align}
    \Psi_{a}(\mathbf{F},\mathnormal{C}) &=
    \frac{\mathnormal{C}\mathnormal{J}}{\mathnormal{C}_{0}}\eta_{\alpha} 
    + \frac{\beta}{2}\left(\frac{\mathnormal{C}}{\mathnormal{C}_{0}}-1\right)^{2} \label{eqn:Psi_bulk_active}
    \\
    \widetilde{\Psi}_{a}(\widetilde{\mathbf{F}},\widetilde{\mathnormal{C}}) &=
    \frac{\widetilde{\mathnormal{C}}\widetilde{\mathnormal{J}}}{\widetilde{\mathnormal{C}}_{0}}\eta_{\widetilde{\alpha}}
    + \frac{\widetilde{\beta}}{2}\left(\frac{\widetilde{\mathnormal{C}}}{\widetilde{\mathnormal{C}}_{0}}-1\right)^{2} \label{eqn:Psi_surf_active}
\end{align}
\end{subequations}
The first terms in \Cref{eqn:Psi_bulk_active,eqn:Psi_surf_active} represent the contribution of cell contractility and concentration to the free energy. Consistent with their definition in our previous work \citep{kim2023model}, the terms $\eta_{\alpha}=\alpha\eta\left(\mathnormal{t}\right)$ and $\eta_{\widetilde{\alpha}}=\widetilde{\alpha}\eta\left(\mathnormal{t}\right)$ correspond to the bulk and surface contractile moduli, respectively, where $\alpha$ and $\widetilde{\alpha}$ are the maximum allowable values modulated by the activation level $\eta\left(\mathnormal{t}\right)$. The activation level of the cell contractility can be viewed as a variable that reflects the stabilization of focal adhesions and stress fibers in the cell cytoskeleton \citep{humphrey2014mechanotransduction,eichinger2021mechanical}, that will be defined in \Cref{sec:Activation level}. Note that $\mathnormal{C}\mathnormal{J}$ and $\widetilde{\mathnormal{C}}\widetilde{\mathnormal{J}}$ are the cell concentrations in the spatial configuration, which are normalized by the initial cell concentrations $\mathnormal{C}_{0}$ and $\widetilde{\mathnormal{C}}_{0}$. The second terms in \Cref{eqn:Psi_bulk_active,eqn:Psi_surf_active} regulate the cell distribution using a penalty approach, where $\beta$ and $\widetilde{\beta}$ are the penalty coefficients, and $\mathnormal{C}/\mathnormal{C}_{0}$ and $\widetilde{\mathnormal{C}}/\widetilde{\mathnormal{C}}_{0}$ are the ratios of cell concentration in the material configuration.

\subsection{Activation level}\label{sec:Activation level}
Cells in living tissues interact with the extracellular matrix (ECM) to maintain homeostasis \citep{brown1998tensional,petroll2004corneal,paszek2005tensional,doyle2009one,eichinger2021mechanical}. To capture the formation of stable focal adhesions between the cells and the ECM, and stable intracellular stress fibers, we have previously introduced a non-dimensional activation level of cell contractility, $\eta\left(\mathnormal{t}\right)$ \citep{kim2023model}, that follows an equilibration process (see \Cref{fig:Fig2}). Living soft tissues exhibit two characteristic stages towards this equilibration process: (phase I) the contractile force increases rapidly, and (phase II) the force reaches a steady-state \citep{delvoye1991measurement,eastwood1996quantitative,brown1998tensional,eichinger2020computer}. Given the complexity of tracking single cell forces, we suggest the following expression that approximately describes the dynamics of activation for all cells in the microtissue,
\begin{align}\label{eqn:Activation level}
    \eta\left(\mathnormal{t}\right) =
    \begin{dcases}
        1-\left(1-\frac{\mathnormal{t}}{\tau}\right)^{\mathnormal{n}} & (t\leq\tau) \\
        1 & (t>\tau)
    \end{dcases} 
\end{align}
In addition, we make the assumption, based on the conformity assumption in Equation \Cref{eqn:bio-chemicalSlavery}, that the activation level $\eta(t)$ is equal to the surface activation level $\widetilde{\eta}(t)$ throughout the system. This assumption implies that the activation level is uniform and consistent across the bulk and surface regions.
\begin{equation}
    \widetilde{\eta}\left(\mathnormal{t}\right) = \eta\left(\mathnormal{t}\right)
\end{equation}
In this context, the activation level $\eta\left(\mathnormal{t}\right)$ is a non-dimensional quantity that tracks the temporal stability of focal adhesions and the cells' ability to exert traction forces, with a range limited to $\eta\left(\mathnormal{t}\right)\in[0,1]$. It should be noted that $\mathnormal{t}$ represents the time elapsed since cell seeding, $\tau$ is the characteristic time for reaching equilibrium, and $\mathnormal{n}$ is a dimensionless parameter. The reference configuration coincides with an initial unreformed state at zero activation level $\eta(\mathnormal{t}=0)=0$, and the rise of $\eta(\mathnormal{t}>0)$ triggers cell contractility and migration that ultimately lead to an equilibrium state. For this work, we assume a quadratic dependence ($\mathnormal{n}=2$) for both active moduli ($\eta_{\alpha}$ and $\eta_{\widetilde{\alpha}}$). Although it is exceedingly complex to track single cell forces, we suggest an approximate expression for the activation dynamics that is applicable to all cells in the microtissue.

\begin{figure*}[!ht]
    \centering
    \begin{subfigure}{0.4\textwidth}
        \centering
        \includegraphics[clip,width=0.95\linewidth]{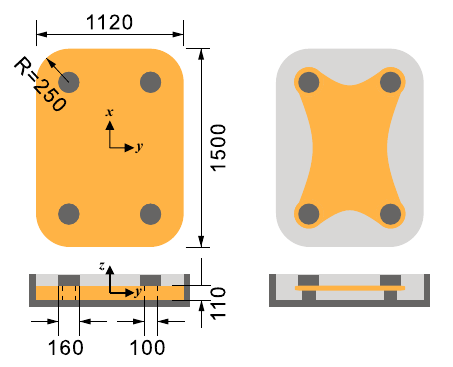} 
        \caption{Unperturbed microtissue}
        \label{fig:Fig3a}
    \end{subfigure}
    \quad
    \begin{subfigure}{0.4\textwidth}
        \centering
        \includegraphics[clip,width=0.95\linewidth]{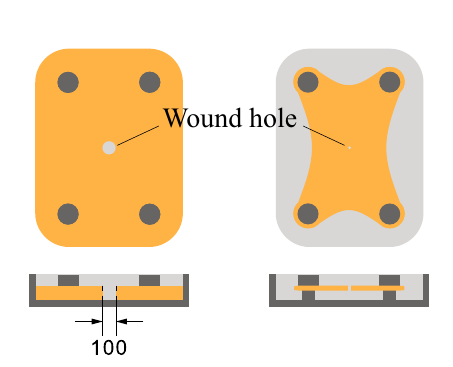}
        \caption{Perturbed microtissue with wound hole}
        \label{fig:Fig3b}
    \end{subfigure}
    \caption{A schematic drawing for (a) an unperturbed microtissue and (b) and a microtissue with a cylindrical wound hole (unit: $\mu\mathrm{m}$). The initial/undeformed (left) and final/equilibrium configurations (right) of microtissues are shown. The sectional view (bottom) is captured at the centerline of the top view (top).}
    \label{fig:Fig3}
\end{figure*}

\subsection{Specific constitutive relations}\label{sec:Specific constitutive relations}
The expression for the first PK stress tensors as well as the cell concentrations are calculated from \Cref{eqn:ConstitutiveRelation} using the free energy in \Cref{eqn:Strain_energy,eqn:Psi_bulk_passive,eqn:Psi_active},
\begin{subequations}\label{eqn:Specific_constitutive_relations}
\begin{align}
    \xi
    &= \frac{\mathnormal{C}}{\mathnormal{C}_{0}}\eta_{\alpha}\mathnormal{J}
    \label{eqn:Specific_constitutive_relations(a)} \\
    \widetilde{\xi}
    &= \frac{\widetilde{\mathnormal{C}}}{\widetilde{\mathnormal{C}}_{0}}\eta_{\widetilde{\alpha}}\widetilde{\mathnormal{J}}
    \label{eqn:Specific_constitutive_relations(b)} \\
    \mathbf{P}
    &= \mathnormal{G}\mathnormal{J}^{-2/3}\left(\mathbf{F}-\frac{1}{3}\mathnormal{I}_{1}\mathbf{F}^{-\mathrm{T}}\right)
    + \mathnormal{K}\left(\mathnormal{J}-1\right)\mathnormal{J}\mathbf{F}^{-\mathrm{T}}
    + \xi\,\mathbf{F}^{-\mathrm{T}} \label{eqn:Specific_constitutive_relations(c)}
    \\
    \widetilde{\mathbf{P}}
    &= \widetilde{\xi}\,\widetilde{\mathbf{F}}^{-\mathrm{T}}  \label{eqn:Specific_constitutive_relations(d)}
    \\
    \mu &= \frac{\beta}{\mathnormal{C}_{0}}\left(\frac{\mathnormal{C}}{\mathnormal{C}_{0}}-1\right) + \frac{1}{\mathnormal{C}_{0}}\eta_{\alpha}\mathnormal{J}
    \Rightarrow 
    \mathnormal{C} = \frac{\mathnormal{C}_{0}}{\beta}\left(\mathnormal{C}_{0}\mu + \beta - \eta_{\alpha}\mathnormal{J}\right) \label{eqn:Specific_constitutive_relations(e)}
    \\
    \widetilde{\mu} &=
    \frac{\widetilde{\beta}}{\widetilde{\mathnormal{C}}_{0}}\left(\frac{\widetilde{\mathnormal{C}}}{\widetilde{\mathnormal{C}}_{0}}-1\right) + \frac{1}{\widetilde{\mathnormal{C}}_{0}}\eta_{\widetilde{\alpha}}\widetilde{\mathnormal{J}}
    \Rightarrow 
    \widetilde{\mathnormal{C}} = \frac{\widetilde{\mathnormal{C}}_{0}}{\widetilde{\beta}}\left(\widetilde{\mathnormal{C}}_{0}\widetilde{\mu}+\widetilde{\beta}-\eta_{\widetilde{\alpha}}\widetilde{\mathnormal{J}}\right) \label{eqn:Specific_constitutive_relations(f)}
\end{align}
\end{subequations}
where $\xi$ and $\widetilde{\xi}$ in \Cref{eqn:Specific_constitutive_relations(a),eqn:Specific_constitutive_relations(b)} are microscopic variables as the energy conjugate of the rate of activation level $\dot{\eta}$, which allows us to capture the accurate representation of the system's behavior. As $\eta$ evolves with time independently of deformation or chemical potential, $\xi$ and $\widetilde{\xi}$ are non-standard variables and are not used in the weak form for finite element implementation. Incorporating the coupled response allows us to capture the sensitivity of the contractile response to both cell concentration and collagen density \citep{eichinger2020computer} as well as the development of forces towards mechanical homeostasis \citep{brown1998tensional}. It is important to note that while the activation level $\eta\left(\mathnormal{t}\right)$ evolves uniformly for all cells, the stresses that develop in the microtissue depend on the coupling of the activation level, the concentration of cells due to migration, and the deformation state. This can lead to an inhomogeneous stress state within the microtissue.

\section{Mixed finite element formulation}\label{sec:Mixed finite element formulation}
This section presents a finite element formulation based on the nonlinear theory described in \Cref{sec:A nonlinear theory,sec:Specific consideration for cell-ECM interaction}. We start with the strong form of the governing equations and initial and boundary conditions. We then introduce the weak form of the problem and subsequently describe the temporal and spatial discretization and solution scheme. As the main focus of this research is the coupled mechanics between cell migration and tissue deformation in bulk and on surface, we neglect the terms associated with proliferation and apoptosis in the bulk and on the surface, i.e., $\mathnormal{r}=\mathnormal{i}=0$, as a simplification.

\begin{figure*}[ht]
    \centering
    \includegraphics[width=0.7\linewidth]{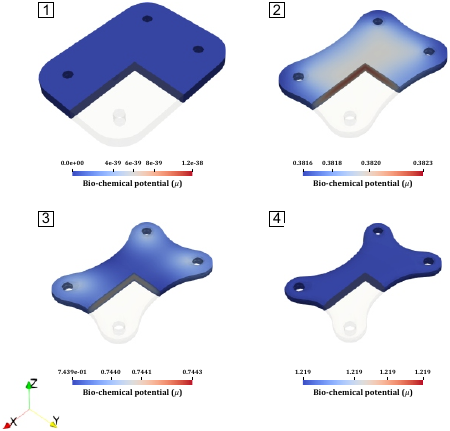}
    \caption{Temporal sequence of the bio-chemical potential during the formation of unperturbed microtissues. A quarter of the domain is made transparent to show the contour plot on two sections of the interior. }
    \label{fig:Fig4}
\end{figure*}

\subsection{Two-field weak form}
To simplify the finite element formulation and avoid the need for $\mathcal{C}^{1}$ continuity requirements \citep{bouklas2015nonlinear}, we use the bio-chemical potential as the independent variable instead of cell concentration. The free energy densities are then expressed as a function of the deformation gradient and bio-chemical potential using a Legendre transform \citep{hong2008theory,mcbride2011geometrically,bouklas2015nonlinear}, which replaces a variable with its thermodynamic conjugate.
\begin{subequations}\label{eqn:TransformedStrainEnergyFunction}
\begin{align}
    \Phi(\mathbf{F},\mu,\eta) &= \Psi(\mathbf{F},\mathnormal{C},\eta) - \mu\mathnormal{C} \\
    \widetilde{\Phi}(\widetilde{\mathbf{F}},\widetilde{\mu},\widetilde{\eta}) &= \widetilde{\Psi}(\widetilde{\mathbf{F}},\widetilde{\mathnormal{C}},\widetilde{\eta}) - \widetilde{\mu}\widetilde{\mathnormal{C}}
\end{align}
\end{subequations}

The weak form of the problem is obtained by using a set of test functions, which satisfy the necessary integrability conditions. By multiplying \Cref{eqn:EquilibriumBulk,eqn:BalanceLawBulk} with $\delta\mathbf{u}$ and $\delta\mu$, and integrating over the domain, respectively, we obtain:
\begin{align}
    \int_{\mathnormal{V}} \mathbf{P}:\boldsymbol{\nabla}_{\mathbf{X}}\delta\mathbf{u} \, \mathrm{d}V
    +\int_{\mathnormal{S}} \widetilde{\mathbf{P}}:\widetilde{\boldsymbol{\nabla}}_{\widetilde{\textbf{X}}}\delta\mathbf{u} \, \mathrm{d}S
    &=0 \label{eqn:WeakForm_Equilibrium} \\
    \int_{\mathnormal{V}} \dot{\mathnormal{C}}\,\delta\mu \, \mathrm{d}V -\int_{\mathnormal{V}} \mathbf{J}\cdot\boldsymbol{\nabla}_{\mathbf{X}}\delta\mu \, \mathrm{d}V
    +  \int_{\mathnormal{S}} \dot{\widetilde{\mathnormal{C}}}\,\delta\mu \, \mathrm{d}S 
    - \int_{\mathnormal{S}} \widetilde{\mathbf{J}}\cdot\widetilde{\boldsymbol{\nabla}}_{\widetilde{\textbf{X}}}\delta\mu\, \mathrm{d}S 
    &= 0 \label{eqn:WeakForm_Balance}
\end{align}
The weak form is a reformulation of the governing equations and boundary conditions that seeks to find trial functions for the displacement field $\mathbf{u}(\mathbf{X},\mathnormal{t})$ and bio-chemical potential $\mu(\mathbf{X},\mathnormal{t})$ that satisfy the equations, when tested against a set of permissible test functions $\delta\mathbf{u}$ and $\delta\mu(\mathbf{X})$. 
In our previous work \citep{kim2023finite} on surface and bulk poroelasticity of hydrogels, we were not able to obtain a closed-form solution for the constitutive law governing the surface concentration. This led us to use a three-field weak form, which required solving an additional nonlinear equation for surface concentration in a weak sense, which led to oscillations near sharp features of the domain. Additionally, we have to note that the three-field weak form inherently has a saddle point structure. In this current work, we choose the specific free-energy formulation that enables us to obtain a two-field weak form without the need to solve an additional nonlinear equation, thereby avoiding numerical oscillations.

\begin{figure*}[ht]
    \centering
    \includegraphics[width=0.7\linewidth]{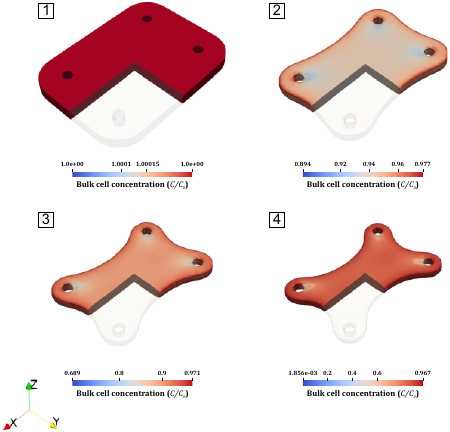}
    \caption{Temporal sequence of the bulk cell concentration during the formation of unperturbed microtissues. A quarter of the domain is made transparent to show the contour plot on two sections of the interior. }
    \label{fig:Fig5}
\end{figure*}

\subsection{Temporal discretization}
The backward Euler scheme is used to integrate \Cref{eqn:WeakForm_Balance} over time:
\begin{align}  \label{eqn:WeakForm_Balance_Backward_Euler}
    &\int_{\mathnormal{V}} \frac{1}{\Delta\mathnormal{t}}\left(\mathnormal{C}^{\mathnormal{t}+\Delta\mathnormal{t}}-\mathnormal{C}^{\mathnormal{t}}\right)\delta\mu \, \mathrm{d}\mathnormal{V} 
    - \int_{\mathnormal{V}} \mathbf{J}^{\mathnormal{t}+\Delta\mathnormal{t}}\cdot\boldsymbol{\nabla}_{\mathbf{X}}\delta\mu \, \mathrm{d}\mathnormal{V}
    \nonumber\\
    &+ \int_{\mathnormal{S}} \frac{1}{\Delta\mathnormal{t}}\left(\widetilde{\mathnormal{C}}^{\mathnormal{t}+\Delta\mathnormal{t}}-\widetilde{\mathnormal{C}}^{\mathnormal{t}}\right)\delta\mu \, \mathrm{d}\mathnormal{S} 
    - \int_{\mathnormal{S}} \widetilde{\mathbf{J}}^{\mathnormal{t}+\Delta\mathnormal{t}}\cdot\widetilde{\boldsymbol{\nabla}}_{\widetilde{\textbf{X}}}\delta\mu\, \mathrm{d}\mathnormal{S}
    = 0
\end{align}
where the superscripts indicate the time step, at the current time step ($\mathnormal{t}+\Delta\mathnormal{t}$) or the previous step $\mathnormal{t}$. We can combine \Cref{eqn:WeakForm_Equilibrium} and \Cref{eqn:WeakForm_Balance_Backward_Euler} as
\begin{align}\label{eqn:WeakForm_Combined}
    &\int_{\mathnormal{V}} \mathbf{P}:\boldsymbol{\nabla}_{\mathbf{X}}\delta\mathbf{u} \, \mathrm{d}\mathnormal{V}
    + \int_{\mathnormal{V}} \left(\mathnormal{C} - \mathnormal{C}^{\mathnormal{t}}\right)\delta\mu\,\mathrm{d}\mathnormal{V} 
    - \Delta\mathnormal{t}\int_{\mathnormal{V}}\mathbf{J}\cdot\boldsymbol{\nabla}_{\mathbf{X}}\delta\mu\, \mathrm{d}\mathnormal{V}
    \nonumber\\
    &+ \int_{\mathnormal{S}} \widetilde{\mathbf{P}}:\widetilde{\boldsymbol{\nabla}}_{\widetilde{\textbf{X}}}\delta\mathbf{u} \, \mathrm{d}\mathnormal{S}  
    + \int_{\mathnormal{S}}\left(\widetilde{\mathnormal{C}} -  \widetilde{\mathnormal{C}}^{\mathnormal{t}}\right)\delta\mu \, \mathrm{d}\mathnormal{S} 
    - \Delta\mathnormal{t}\int_{\mathnormal{S}}\widetilde{\mathbf{J}}\cdot\widetilde{\boldsymbol{\nabla}}_{\widetilde{\textbf{X}}}\delta\mu\, \mathrm{d}\mathnormal{S}
    = 0
\end{align}
where the superscript ($\mathnormal{t}+\Delta\mathnormal{t}$) is omitted for all the terms at the current time step and $\mathnormal{C}^{\mathnormal{t}}$ and $\widetilde{\mathnormal{C}}^{\mathnormal{t}}$ are the species concentration at the previous time step in the bulk and on the surface.

\begin{figure*}[ht]
    \centering
    \includegraphics[width=0.7\linewidth]{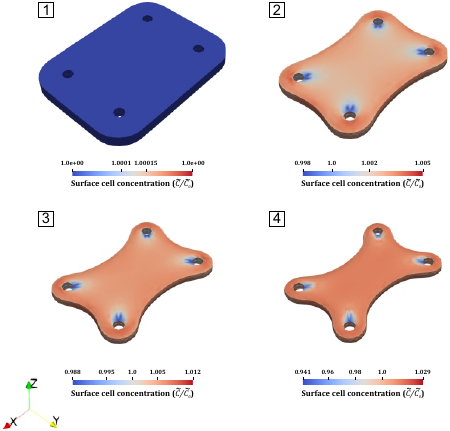}
    \caption{Temporal sequence of the surface cell concentration during the formation of unperturbed microtissues.}
    \label{fig:Fig6}
\end{figure*}

\subsection{Spatial discretization}
To solve for the displacement and bio-chemical potential fields simultaneously, a mixed finite element method is utilized. However, the mixed method can be numerically unstable if proper spatial discretization techniques are not employed \citep{bouklas2015nonlinear}. To ensure numerical stability, an equal-order interpolation for displacement and bio-chemical potential was first tested, but it led to spurious oscillations in the solution. To overcome this issue, the MINI element \citep{arnold1984stable} was implemented, which uses equal-order interpolations for displacement and bio-chemical potential but enriches the interpolation for displacement with a bubble function \citep{boffi2013mixed}. The displacement and bio-chemical potential are then interpolated throughout the domain of interest using the MINI element.
\begin{equation}\label{eqn:Spatial_Discretization_trial_function}
    \mathbf{u} = \mathbf{H}^{\mathbf{u}}\mathbf{u}^{n}
    \quad \text{and} \quad
    \mu = \mathbf{H}^{\mu}\boldsymbol{\mu}^{n}
\end{equation}
where $\mathbf{H}^{\mathbf{u}}$ and $\mathbf{H}^{\mu}$ are the shape functions, $\mathbf{u}^{n}$ and $\boldsymbol{\mu}^{n}$ are the nodal values of the displacement and bio-chemical potential, respectively. The test functions are discretized in the same way
\begin{equation}\label{eqn:Spatial_Discretization_test_function}
    \delta\mathbf{u} = \mathbf{H}^{\mathbf{u}}\delta\mathbf{u}^{n}
    \quad \text{and} \quad
    \delta\mu = \mathbf{H}^{\mu}\delta\boldsymbol{\mu}^{n}
\end{equation}
The stress, concentration, and flux are evaluated at integration points, depending on the gradients of the displacement and bio-chemical potential via the constitutive relations. Taking the gradient of \Cref{eqn:Spatial_Discretization_test_function}, we obtain that
\begin{subequations}
\begin{alignat}{2}
    \boldsymbol{\nabla}_{\mathbf{X}}\delta\mathbf{u} &= \boldsymbol{\nabla}_{\mathbf{X}}\mathbf{H}^{\mathbf{u}}\delta\mathbf{u}^{n} &&= \mathbf{B}^{\mathbf{u}}\delta\mathbf{u}^{n} \\
    \widetilde{\boldsymbol{\nabla}}_{\widetilde{\textbf{X}}}\delta\mathbf{u} &= \widetilde{\boldsymbol{\nabla}}_{\widetilde{\textbf{X}}}\mathbf{H}^{\mathbf{u}}\delta\mathbf{u}^{n} &&= \widetilde{\mathbf{B}}^{\mathbf{u}}\delta\mathbf{u}^{n} \\
    \boldsymbol{\nabla}_{\mathbf{X}}\delta\mu &= \boldsymbol{\nabla}_{\mathbf{X}}\mathbf{H}^{\mu}\delta\boldsymbol{\mu}^{n} &&= \mathbf{B}^{\mu}\delta\boldsymbol{\mu}^{n} \\
    \widetilde{\boldsymbol{\nabla}}_{\widetilde{\textbf{X}}}\delta\mu &= \widetilde{\boldsymbol{\nabla}}_{\widetilde{\textbf{X}}}\mathbf{H}^{\mu}\delta\boldsymbol{\mu}^{n} &&= \widetilde{\mathbf{B}}^{\mu}\delta\boldsymbol{\mu}^{n}
\end{alignat}
\end{subequations}
where $\mathbf{B}^{\mathbf{u}}$ and $\mathbf{B}^{\mu}$ are the gradients of the shape functions in the bulk, and $\widetilde{\mathbf{B}}^{\mathbf{u}}$ and $\widetilde{\mathbf{B}}^{\mu}$ are the ones on the surface.

\begin{figure*}[ht]
    \centering
    \includegraphics[width=0.5\linewidth]{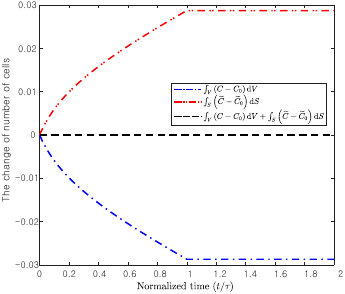}
    \caption{Contractile forces drive cells to move from the interior to the surface of the extracellular matrix (ECM). We track the number of species in both the bulk and on the surface over time. During the activation level ramping phase ($\mathnormal{t}/\tau < 1.0$), we observe an increase in the number of cells on surface and a decrease in bulk, while the total number of species remains constant. Once the activation level is fixed ($\mathnormal{t}/\tau \geq 1.0$), cell migration is suspended.}
    \label{fig:Fig7}
\end{figure*}

\subsection{Nonlinear solution}

The weak form in \Cref{eqn:WeakForm_Combined} can be expressed as a system of nonlinear equations,
\begin{equation}\label{eqn:NonlinearEquation}
    \mathcal{N}(\mathbf{d}) = \mathbf{f} \quad \text{with} \quad \mathbf{d}
    = \left[ \mathbf{u}^{n} 
    \;  
    \boldsymbol{\mu}^{n}
    \right]^{\mathrm{T}}
\end{equation}
We note that $\mathcal{N}(\mathbf{d})$ represents the unknown part of the weak form at the current time step. To obtain a solution, we move all the known quantities to the right-hand side and denote them as $\mathbf{f}$, which is given from the previous time step. The residual of the nonlinear equations at iteration step $\mathnormal{i}$ is then given by $\mathbf{R}_{\mathnormal{i}}=\mathbf{f}-\mathcal{N}(\mathbf{d}_{\mathnormal{i}})$, which can be solved using the Newton-Raphson method. The method involves the calculation of the tangent Jacobian matrix at each iteration step, which is given by:
\begin{equation}\label{eqn:BlockJacobianMatrix}
    \frac{\partial\mathcal{N}}{\partial\mathbf{d}}\bigg\vert_{\mathbf{d}_{i}} 
    = \begin{bmatrix}
    \mathbf{K}^{\mathbf{u}\mathbf{u}} 
    & \mathbf{K}^{\mathbf{u}\mu} \\
    \mathbf{K}^{\mu\mathbf{u}} 
    & \mathbf{K}^{\mu\mu}
    \end{bmatrix}
\end{equation}
The coupled non-linear equations are numerically solved using FEniCS version 2019.2.0 \citep{LoggMardalEtAl2012a,AlnaesBlechta2015a}, Multiphenics \citep{ballarin2019multiphenics} and the Portable Extensible Toolkit for Scientific Computations (PETSc) Scalable Nonlinear Equations Solvers (SNES) interface \citep{balay2019petsc}. The solution process continues until a specified level of convergence is achieved within the SNES solver.

\begin{figure*}[ht]
    \centering
    \includegraphics[width=0.7\linewidth]{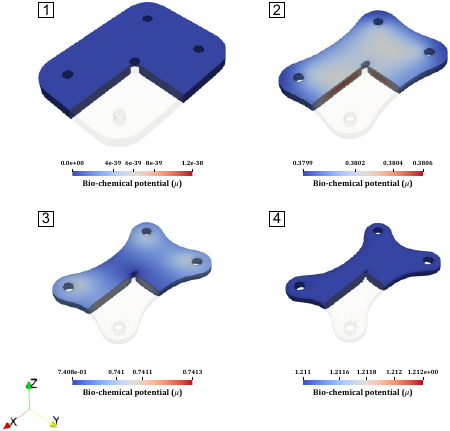}
    \caption{Temporal sequence of the bio-chemical potential during the formation of perturbed microtissues. A quarter of the domain is made transparent to show the contour plot on two sections of the interior.}
    \label{fig:Fig8}
\end{figure*}

\section{Results and discussion}\label{sec:Results and discussion}
To test the robustness of our numerical approach, and also the predictions corresponding to the specific model choices, we simulated two boundary value problems that we have studied experimentally in our previous work \citep{mailand2019surface, sakar2016cellular}. The first problem involves simulating the equilibrium morphology and cell distribution of microtissues during unperturbed morphogenesis in a microengineered \emph{in vitro} platform. The second problem refers to simulating the equilibrium morphology and cell distribution of microtissues that are surgically perturbed using a robotic micromanipulation system. \Cref{fig:Fig3} shows a schematic illustration of microtissues in their initial and final configurations for the two distinct cases. In the second case, a cylindrical hole is prescribed in the center of the reference state (see \Cref{fig:Fig3b}, left). 

\begin{figure*}[ht]
    \centering
    \includegraphics[width=0.7\linewidth]{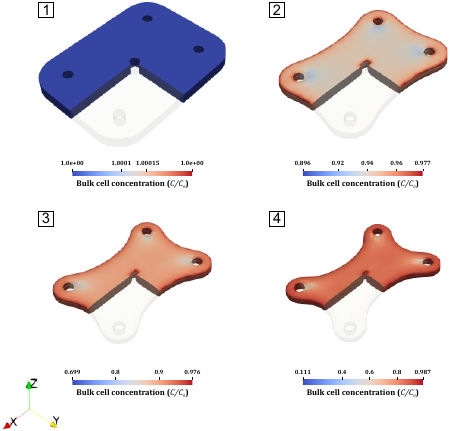}
    \caption{Temporal sequence of the bulk cell concentration during the formation of perturbed microtissues. A quarter of the domain is made transparent to show the contour plot on two sections of the interior.}
    \label{fig:Fig9}
\end{figure*}

The boundary conditions for the two cases are identical, where the contact surfaces with the pillars are assumed to slide in the axial direction ($\textit{z}$-direction) in a frictionless manner. The rest of the surface is considered to be traction-free, and rigid body motions are prevented. The initial normalized cell concentration is set to unity ($\mathnormal{C}/\mathnormal{C}_{0}=\widetilde{\mathnormal{C}}/\widetilde{\mathnormal{C}}_{0}=1$) at the beginning of the simulation ($\mathnormal{t}/\tau=0$), which determines the initial bio-chemical potentials ($\mu_{0}=\widetilde{\mu}_{0}=0$) through equations \Cref{eqn:Specific_constitutive_relations(e)} and \Cref{eqn:Specific_constitutive_relations(f)}. The elastic modulus and Poisson’s ratio are taken as $\mathnormal{E}=17.38\mathrm{kPa}$ and $\nu=0.09$, respectively, resulting in bulk and shear moduli of $\mathnormal{K}=7\mathrm{kPa}$ and $\mathnormal{G}=8\mathrm{kPa}$ \citep{kim2020model,kim2023model}. The active moduli, $\alpha=25\mathrm{kPa}$, $\widetilde{\alpha}=2\mathrm{mN}/\mathrm{mm}$, $\beta=50\mathrm{kPa}$, and $\widetilde{\beta}=40\mathrm{mN}/\mathrm{mm}$, are calibrated using empirical characterization of the final tissue shape. The activation level of contractility is set to $\mathnormal{n}=2$, and the time-normalized effective diffusivity was selected as $\mathnormal{D}\tau = \widetilde{\mathnormal{D}}\tau = 1.0 \mathrm{m}^{2}\mathrm{J}^{-1}$. The characteristic time is chosen as $\tau=24\mathrm{hr}$ based on observations that the tissues contracted at a significantly lower rate after the first 24 hours in previous experimental studies \citep{mailand2019surface,kim2020model}. The activation level of the cell contractility $\eta\left(\mathnormal{t}\right)$ drives the response in the simulations, while no external mechanical force is applied to the system. To maintain numerical stability, the activation level $\eta\left(\mathnormal{t}\right)$ is increased gradually from zero to its prescribed value for the time interval $\mathnormal{t}/\tau \in [0.0,1.0]$, and then the time steps $\Delta\mathnormal{t}/\tau$ are exponentially increased until equilibrium is attained with the active moduli being fixed.

\begin{figure*}[ht]
    \centering
    \includegraphics[width=0.7\linewidth]{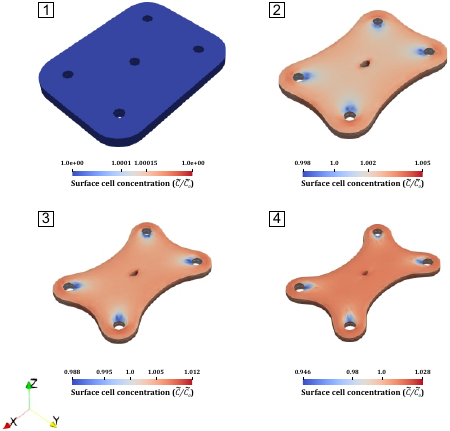}
    \caption{Temporal sequence of the surface cell concentration during the formation of perturbed microtissues.}
    \label{fig:Fig10}
\end{figure*}

\subsection{Time-dependent cell contractility and migration during tissue formation}\label{sec:Time-dependent cell contractility and migration in unperturbed microtissues}
In this section, we explore the nonlinear and transient response of microtissues during the uninterrupted morphogenesis process (see \Cref{fig:Fig3a}). We plot the temporal sequence of the finite element simulation for the bio-chemical potential, bulk concentration, and surface concentration in \Cref{fig:Fig4,fig:Fig5,fig:Fig6}, respectively, with images that are taken at normalized time $\mathnormal{t}/\tau=0.0$, $0.2$, $0.5$, $1.0$ (indicated by Step 1, 2, 3 and 4). In the initial stress-free undeformed state, corresponding to a cell-laden crosslinked hydrogel, the active free energies $\Psi_{a}$ and $\widetilde{\Psi}_{a}$ are taken as zero because we assume that the cells have not started interacting with the ECM yet (Step 1). Through the prescribed evolution law of $\eta\left(\mathnormal{t}\right)$ in \Cref{eqn:Activation level}, the activation level is gradually increased from zero to unity as a function of time $\mathnormal{t}$, leading to microtissue contraction and cell migration. At low activation ($\eta=0.2$), even at an early stage $\mathnormal{t}/\tau=0.2$, the deformation is noticeable, and the mechanosensitive mechanisms captured by the proposed model guide the cells to the periphery of the microtissue and regions near the pillars (Step 2). Beyond $\mathnormal{t}/\tau=0.5$, minimal migration and contractility are observed (Step 3). Finally, increased cell concentration near the periphery generates higher contraction, leading to smoother concave edges on the sides of the microtissue (Step 4).

\begin{figure*}[ht]
    \centering
    \begin{subfigure}[b]{0.4\textwidth}
        \centering
        \includegraphics[width=\textwidth]{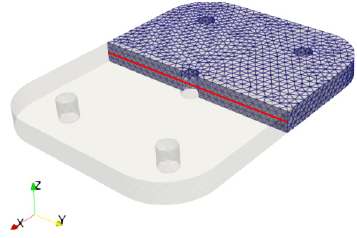}
        \caption{The nodes on the red line are to be tracked over time.}
        \label{fig:Fig11a}
    \end{subfigure}
    \quad 
    \begin{subfigure}[b]{0.4\textwidth}  
        \centering 
        \includegraphics[width=\textwidth]{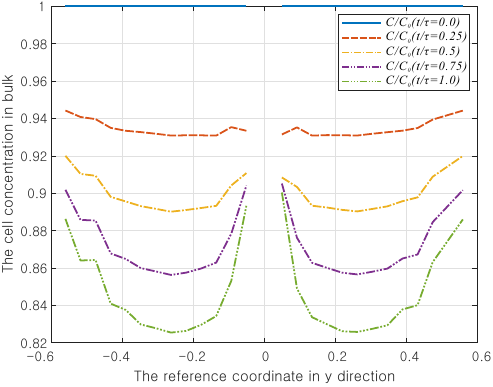}
        \caption{The evolution of the bulk cell concentration on the red line.}    
        \label{fig:Fig11b}
    \end{subfigure}
    \caption{During the activation level ramping phase, we obtain the bulk cell concentration of the perturbed microtissue along the centerline. We observe an overall decrease in bulk cell concentration, but there is an increase in concentration near the wound hole and microtissue boundaries as tissue formation progresses.}
    \label{fig:Fig11}
\end{figure*}

In contrast to our previous theories \citep{kim2020model,kim2023model} where we assumed that the cell concentrations coincide in bulk and on the surface, i.e., $\mathnormal{C}\vert_{\mathnormal{S}}=\widetilde{\mathnormal{C}}$, we can account for the distinct response here, as shown in \Cref{fig:Fig5,fig:Fig6}. On top of cell migration in bulk and on the surface, there is an exchange of cells between the surface and the bulk, as captured in the strong form of the balance law in \Cref{eqn:BalanceLaw}. Cell migration between bulk and surface can be tracked by integrating the normalized concentrations over the bulk and the surface at all time steps, where we observe cell migration from bulk to the surface while the total number of cells is conserved (\Cref{fig:Fig7}). To the best of the authors' knowledge, this is the first time such multiphysical biological processes have been studied in a macroscale simulation.

\subsection{Time-dependent cell contractility and migration during wound closure}
In this section, we simulate a microtissue model with a hole of in the center (see \Cref{fig:Fig3b}). All other dimensions, material properties, boundary conditions, as well as the ramping profile of activation are the same as in the previous example described in \Cref{sec:Time-dependent cell contractility and migration in unperturbed microtissues}. The temporal sequence of the finite element simulations for the bio-chemical potential, bulk concentration, surface concentration are shown in \Cref{fig:Fig8,fig:Fig9,fig:Fig10}, respectively, with images that are taken at normalized time $\mathnormal{t}/\tau=0.0$, $0.2$, $0.5$, $1.0$ (indicated by Step 1, 2, 3 and 4). When the activation level is set to zero ($\eta=0$), the perturbed microtissue is considered to be in a stress-free undeformed state with a hole in the center (Step 1). Through the prescribed evolution law of $\eta\left(\mathnormal{t}\right)$ in \Cref{eqn:Activation level}, the activation level is gradually increased from zero to unity as a function of time $\mathnormal{t}$, leading to wound closure. At an early stage $\mathnormal{t}/\tau=0.2$, the mechanosensitive mechanisms captured by the model guide the cells to the periphery of the wound hole and the outer boundary of microtissue (Step 2). Beyond $\mathnormal{t}/\tau=0.5$, minimal migration and contractility are observed (Step 3). Finally, The cell migration and contractility result in effectively closing the wound hole, which we call the equilibrium state of perturbed microtissue (Step 4).

\begin{figure*}[ht]
    \centering
    \begin{subfigure}[b]{0.4\textwidth}
        \centering
        \includegraphics[width=\textwidth]{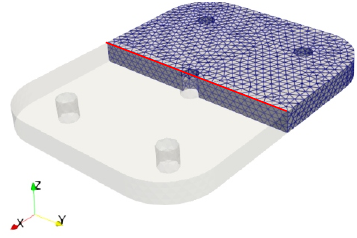}
        \caption{The nodes on the red line are to be tracked over time.}
        \label{fig:Fig12a}
    \end{subfigure}
    \quad 
    \begin{subfigure}[b]{0.4\textwidth}  
        \centering 
        \includegraphics[width=\textwidth]{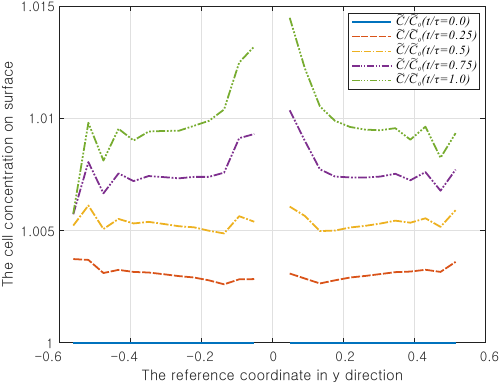}
        \caption{The evolution of the surface cell concentration on the red line.} 
        \label{fig:Fig12b}
    \end{subfigure}
    \caption{During the activation level ramping phase, we obtain the surface cell concentration of the perturbed microtissue along the centerline. Our results show an overall increase in surface cell concentration, with a locally higher concentration near the wound hole.}
    \label{fig:Fig12}
\end{figure*}

\begin{figure*}[ht]
    \centering
    \includegraphics[width=0.5\linewidth]{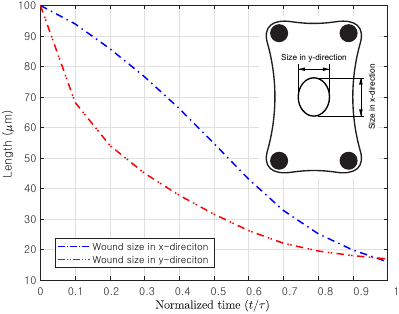}
    \caption{The proposed model successfully captures the wound closure process, as demonstrated by measuring the wound hole in two orthogonal directions. The results show that as the activation level ramps up, the wound gradually closes.}
    \label{fig:Fig13}
\end{figure*}

We further investigated the evolution of cell concentration of the perturbed microtissues to account for the distinct response of cells migration in bulk and on surface. We set the centerline at the middle of the bulk and on top of the surface, as shown in \Cref{fig:Fig11a,fig:Fig12a}, respectively. The profiles of cell concentration along the centerlines are reported at the normalized time $\mathnormal{t}/\tau=0.0$, $0.25$, $0.5$, $0.75$, $1.0$ with respect to the reference position. We can understand the local trend of cell migration by observing the decrease in bulk (\Cref{fig:Fig11b}) and the increase on surface (\Cref{fig:Fig12b}), which is consistent with the global trend of cell migration depicted in \Cref{fig:Fig7}. Furthermore, the results reveal the heterogeneous nature of cell migration along the centerline as the cell concentration at the boundaries of the microtissue and wound periphery are greater than those of the interior. The elevated cell concentration increases the surface energy, ultimately contributing to the closure of the wound. These results are consistent with the experimental observations reported in \citet{sakar2016cellular}. Fibroblasts move into the boundaries of the gap during the closure of the wound. Finally, we plot the size of the wound hole in both the $\textit{x}$ and $\textit{y}$ directions over normalized time (\Cref{fig:Fig13}), and we observe a monotonic reduction in wound size, around 16 $\mu\mathrm{m}$, consistent with our experimental results \citep{sakar2016cellular}. 

\section{Conclusion}
We introduce a continuum mechanics framework and corresponding open-source finite element implementation that considers the interaction between cellular contractility, migration, and ECM mechanics in dynamically morphing soft tissues. The proposed multiphysics model incorporates surface and bulk contractile stresses as well as surface and bulk cell kinetics driven by mechanosensing, and a direct coupling between the local deformation state and contractile force generation. Simulation results demonstrate the potential of our model to capture changes in tissue shape and cell concentration for both intact and mechanically manipulated microtissues. Such studies can provide valuable insights on cell-ECM interactions in the context of development, oncology, tissue engineering, and regenerative medicine.

This study reveals that cell migration in microtissues is influenced by the local geometry, with cells migrating towards the outer boundaries of the tissue and wound periphery. This directed migration is likely driven by the higher surface energy at the boundaries, which helps maintain the geometric integrity of the microtissues. During the early stages of morphogenesis, cells inside the microtissues exhibit high mobility and rapidly migrate towards the wound periphery. Notably, the migration pattern changes over time. In the early stage of cell migration ($\mathnormal{t}/\tau<0.2$), we find that the initial cell migration is primarily determined by the geometry, as illustrated in \Cref{fig:Fig5,fig:Fig6,fig:Fig9,fig:Fig10} Step 2. However, in late stages ($\mathnormal{t}/\tau>0.5$), the geometry of the microtissue no longer has a significant effect, and the cell migration maintains the previous direction, as shown in \Cref{fig:Fig5,fig:Fig6,fig:Fig9,fig:Fig10} Steps 3 and 4. In our future work, we will experimentally test these hypothesis by tracking cells using time-lapse imaging.

Taken together, our results suggest that local geometry is an important contributing factor that organizes cell migration in microtissues. From a computational standpoint, our approach is robust as we test multiple challenging and highly nonlinear boundary value problems with unique features. Interestingly, our implementation, which is based on a two-field weak form, overcomes numerical oscillation issues that arise near sharp features of the geometry in a three-field weak form-based implementation for surface and bulk poroelasticity of hydrogels, as described in our previous work \citep{kim2023finite}. 

\section*{Declaration of Competing Interest}
The authors declare that they have no known competing financial interests or personal relationships that could have appeared to influence the work reported in this paper.

\section*{Acknowledgments}
The authors thank prof. Berkin Dortdivanlioglu for pointing us to the work of \citet{lucantonio2016continuum}.
\appendix
\section{Thermodynamics considerations}\label{sec:Thermodynamics considerations}

Considering a system that includes an elastic collagen network coupled with cells that are free to migrate, its free energy, $\mathcal{G}$, has to account for these energetic contributions \citep{holzapfel2000nonlinear,gurtin2010mechanics,hong2008theory,bouklas2015nonlinear,ang2020effect, mao2018theory}.
\begin{equation}\label{eqn:FreeEnergyOfSystem}
    \dot{\mathcal{G}}
    = \int_{\mathnormal{V}} \dot{\Psi} \,\mathrm{d}V
    + \int_{\mathnormal{S}} \dot{\widetilde{\Psi}} \,\mathrm{d}S
    - \int_{\mathnormal{V}} \mathbf{B}\dot{\mathbf{x}} \,\mathrm{d}V
    - \int_{\mathnormal{S}} \mathbf{T}\dot{\mathbf{x}} \,\mathrm{d}S
    - \int_{\mathnormal{V}} \mu\mathnormal{r} \,\mathrm{d}V
    - \int_{\mathnormal{S}} \widetilde{\mu}\mathnormal{i} \,\mathrm{d}S
    - \int_{\mathnormal{V}} \xi\dot{\eta} \,\mathrm{d}V
    - \int_{\mathnormal{S}} \widetilde{\xi}\dot{\widetilde{\eta}} \,\mathrm{d}S
\end{equation}
where the third and fourth terms are the rate of mechanical work, which can be obtained by integrating \Cref{eqn:Equilibrium}. The fifth and sixth terms are the rate of bio-chemical work, which can be obtained by integrating \Cref{eqn:BalanceLaw}. The seventh and eighth terms in the equation represent the microscopic force system, which is non-standard \citep{mao2018theory}. In this system, the microforces $\xi$ and $\widetilde{\xi}$ in the bulk and on the surface, respectively, are the energy conjugates of the rate of activation levels $\dot{\eta}$ and $\dot{\widetilde{\eta}}$. Note that $\dot{\{\bullet\}}$ is the time derivative of the quantity, and the thermodynamics dictate that the free energy of the system should not increase, i.e., $\dot{\mathcal{G}}\leq0$.

Consequently, coupling species diffusion and polymer deformation problem, the rate of change of the free energy of the system can be expressed as follows,
\begin{align}
    \dot{\mathcal{G}}
    &= \int_{\mathnormal{V}} \dot{\Psi} \,\mathrm{d}V
    +\int_{\mathnormal{S}} \dot{\widetilde{\Psi}} \,\mathrm{d}S
    - \int_{\mathnormal{V}} \mathbf{P}:\dot{\mathbf{F}} \,\mathrm{d}V - \int_{\mathnormal{S}}  \widetilde{\mathbf{P}}:\dot{\widetilde{\mathbf{F}}} \,\mathrm{d}S
    - \int_{\mathnormal{V}} \mu\dot{\mathnormal{C}} \,\mathrm{d}V 
    - \int_{\mathnormal{S}} \widetilde{\mu}\dot{\widetilde{\mathnormal{C}}} \,\mathrm{d}S
    \nonumber\\
    &- \int_{\mathnormal{V}} \xi\dot{\eta} \,\mathrm{d}V
    - \int_{\mathnormal{S}} \widetilde{\xi}\dot{\widetilde{\eta}} \,\mathrm{d}S
    + \int_{\mathnormal{S}} \left(\widetilde{\mu}-\mu\right) \mathbf{J}\cdot\mathbf{N} \,\mathrm{d}S
    + \int_{\mathnormal{V}} \mathbf{J}\cdot\boldsymbol{\nabla}_{\mathbf{X}}\mu \,\mathrm{d}V
    + \int_{\mathnormal{S}} \widetilde{\mathbf{J}}\cdot\widetilde{\boldsymbol{\nabla}}_{\widetilde{\textbf{X}}}\widetilde{\mu} \,\mathrm{d}S
    \leq 0 \label{eqn:DissipationInequality}
\end{align}
Using the chain-rule, the rate of free energy densities can be written by
\begin{equation}\label{eqn:StrainEnnergyChainRule}
    \dot{\Psi} 
    = \frac{\partial\Psi}{\partial\mathbf{F}}:\dot{\mathbf{F}} + \frac{\partial\Psi}{\partial\mathnormal{C}}\dot{\mathnormal{C}}
    + \frac{\partial\Psi}{\partial\eta}\dot{\eta}
    \quad \text{and} \quad
    \dot{\widetilde{\Psi}}
    = \frac{\partial\widetilde{\Psi}}{\partial\widetilde{\mathbf{F}}}:\dot{\widetilde{\mathbf{F}}} 
    + \frac{\partial\widetilde{\Psi}}{\partial\widetilde{\mathnormal{C}}}\dot{\widetilde{\mathnormal{C}}}
    + \frac{\partial\widetilde{\Psi}}{\partial\widetilde{\eta}}\dot{\widetilde{\eta}}
\end{equation}
By substituting \Cref{eqn:StrainEnnergyChainRule} into \Cref{eqn:DissipationInequality}, and rearranging terms yields
\begin{align}\label{eqn:DissipationInequalityChainRule}
    \dot{\mathcal{G}}
    &= \int_{\mathnormal{V}} \left(\frac{\partial\Psi}{\partial\mathbf{F}}-\mathbf{P}\right):\dot{\mathbf{F}} \,\mathrm{d}V
    + \int_{\mathnormal{S}} \left(\frac{\partial\widetilde{\Psi}}{\partial\widetilde{\mathbf{F}}}-\widetilde{\mathbf{P}}\right):\dot{\widetilde{\mathbf{F}}} \,\mathrm{d}S 
    + \int_{\mathnormal{V}} \left(\frac{\partial\Psi}{\partial\mathnormal{C}}-\mu\right)\dot{\mathnormal{C}} \,\mathrm{d}V 
    + \int_{\mathnormal{S}} \left(\frac{\partial\widetilde{\Psi}}{\partial\widetilde{\mathnormal{C}}}-\widetilde{\mu}\right)\dot{\widetilde{\mathnormal{C}}} \,\mathrm{d}S \nonumber\\
    &    + \int_{\mathnormal{V}} \left(\frac{\partial\Psi}{\partial\eta}-\xi\right)\dot{\eta} \,\mathrm{d}V 
    + \int_{\mathnormal{S}} \left(\frac{\partial\widetilde{\Psi}}{\partial\widetilde{\eta}}-\widetilde{\xi}\right)\dot{\widetilde{\eta}} \,\mathrm{d}S 
    + \int_{\mathnormal{S}} \left(\widetilde{\mu}-\mu\right) \mathbf{J}\cdot\mathbf{N} \,\mathrm{d}S 
    + \int_{\mathnormal{V}} \mathbf{J}\cdot\boldsymbol{\nabla}_{\mathbf{X}}\mu \,\mathrm{d}V
    + \int_{\mathnormal{S}} \widetilde{\mathbf{J}}\cdot\widetilde{\boldsymbol{\nabla}}_{\widetilde{\textbf{X}}}\widetilde{\mu} \,\mathrm{d}S
    \leq 0
\end{align}
where each integral represents a distinct mechanism of energy dissipation, associated with mechanical and bio-chemical, and microscopic works. The inequality must hold at every point of the continuum body and for all times during a thermodynamic process. Therefore, every individual integrand in \Cref{eqn:DissipationInequalityChainRule} must either be negative or vanish.

\bibliographystyle{unsrtnat}
\bibliography{references}

\end{document}